\documentclass[runningheads]{llncs}

\usepackage{kotex}
\usepackage{amsmath}
\usepackage[normalem]{ulem}
\useunder{\uline}{\ul}{}
\usepackage{multirow}
\usepackage{booktabs} % For pretty tables
\usepackage{caption} % For caption spacing
\usepackage{subcaption} % For sub-figures
\usepackage{graphicx}
\usepackage{pgfplots}
\usepackage[all]{nowidow}
\usepackage[utf8]{inputenc}
\usepackage{tikz}
\usetikzlibrary{er,positioning,bayesnet}
\usepackage{multicol}
\usepackage{algpseudocode,algorithm,algorithmicx}
\usepackage[inline]{enumitem} % Horizontal lists

\usepackage{geometry}
\definecolor{blue}{HTML}{1F77B4}
\definecolor{orange}{HTML}{FF7F0E}
\definecolor{green}{HTML}{2CA02C}

\pgfplotsset{compat=1.14}

\definecolor{c}{rgb}{0,0.6,0.6}
\definecolor{m}{rgb}{0.6,0,0.6}
\definecolor{y}{rgb}{0.6,0.6,0}
\definecolor{dkred}{rgb}{0.6,0,0}
\definecolor{dkgreen}{rgb}{0,0.6,0}
\definecolor{dkblue}{rgb}{0,0,0.6}

\begin{document}
\title{Crime and social environments:\\Differences between misdemeanors and felonies}

%나중에 고치기
\author{Juyoung Kim\inst{1} \and
Jinhyuk Yun\inst{1, *}}

\institute{School of AI Convergence, Soongsil University, Seoul 06978, South Korea\\
*Corresponding author: \email{jinhyuk.yun@ssu.ac.kr}}
\maketitle              
\begin{abstract}
Owing to the growing population density of urban areas, many people are being increasingly exposed to criminal activity. Increasing crime rates raise the risk of both physical and psychological injury to law-abiding citizens, creating anxiety. From the viewpoint of complex systems, crime prevention through data science can be a solution to such issues. However, previous studies have focused only on a single aspect of crime, ignoring the complex interplay between the various characteristics, which must be considered in an analysis to understand the dynamics underlying criminal activities. In this study, we examined 12 features that have been identified as correlates of crime rates using state-level statistics from the USA. We found that the correlates of misdemeanors and felonies differ. The number of misdemeanors is strongly associated with the police precinct, whereas felony rates are strongly correlated with gun possession and happiness. Our findings suggest that the countermeasures for misdemeanors should be treated differently from those for felonies.
\end{abstract}

\section{Introduction}
As a result of urbanization, people are now more likely to be exposed to crime on a daily basis~\cite{shelley1981crime}. Witnessing crime in one's everyday life is unpleasant, essentially resulting in fear ~\cite{Hale1996Fear}. With an increasing crime rate, people might worry that they will also become a target of crime if they do not receive timely protection from government authorities. Although simple solutions are available, such as surveillance cameras~\cite{Piza2019CCTV} or self-defense weapons, such approaches are essentially temporary and thus cannot address the fundamental societal fear of crime. Therefore, the ultimate goal is to reduce criminal activity. In this study, we examined the correlates of crime to allow people to seek the underlying factors of crime rates. For this purpose, we investigated 12 features that were previously identified as significant correlates of crime rates with crime and social statistics at the state level in the USA ~\cite{kyklos2010Does,Phillips2006relationship,Field1992effect,Lindstr2013More,Heidensohn1989Gender,Lochner2020Education,Tsebelis1990Penalty,Nikolic2014Making,Watts1931influence,Siegel2013relationship,Boakye2010Studying}. In many previous studies, these correlates were treated separately in each context; however, we believe that an entire set of correlates must be encompassed at once to completely illustrate the complexity of criminal activities. Furthermore, minor crimes, namely misdemeanors, are commonly neglected in such analyses. However, disorder and incivility within a community can also result in serious crimes (felonies)~\cite{wilson1982broken}. In this study, we conducted a comparative analysis between misdemeanors and felonies to explore potential ways to reduce crime rates.

There are various types of criminal offenses. For example, such offenses may be classified as crimes against a person or property, or as corporate, statutory, financial, or other types of crimes. In the USA and other common-law jurisdictions, a felony is defined as a serious crime, whereas a misdemeanor is considered less serious. To investigate the differences between these two main types of crimes, we employed state-level crime rate statistics from the USA. Unfortunately, the detailed classification between misdemeanors and felonies differs by state. For instance, whereas drug possession is a felony in some states (\textit{e.g.}, Arizona and Ohio), it is a misdemeanor in other states (\textit{e.g.}, Colorado and California). Despite such differences, state laws must adhere to the federal constitution; thus, the different laws are similar in practical terms, and their statistics can be a good candidate for examining the distinction between misdemeanors and felonies. To determine whether perpetrators of misdemeanors truly behave differently than those of felonies, we begin with the correlation between each type of crime and 12 socioeconomic statuses for each state. Using a regression analysis, we demonstrate that the major correlates of felonies and a misdemeanors are significantly different. To corroborate our observations, we also computed the importance of each factor using modern machine learning algorithms.

The remainder of this paper is organized as follows. Section II introduces the criminal data and methodology used in this investigation. In Section III, we show that the primary correlates of misdemeanors and felonies are distinct based on two types of analysis: i) linear regression and ii) tree-based ensemble model regression. Finally, in Section IV, we provide some concluding remarks and a discussion of future research.

\section{Methods}

\subsection{Criminal datasets}
All countries have a unique judicial system and define crimes differently. Consequently, it is difficult to analyze crime data on a global scale. In addition, in the case of South Korea, the disparity in economic and social indicators between regions is slight, and the degree of exchange between regions is high, yielding a comparison of regional differences inappropriate. To overcome such difficulties, we used state-level crime data from the USA. Each state has its own unique legal system, and the social climate varies considerably between states. There is also a distinction between federal and state laws; however, all laws should be consistent with the legal principles of the USA as a whole. Therefore, state-specific data for the USA may be estimated as an adequate replacement for global data. First, we gathered the number of crimes committed in each state in 2018, as originally collected by the Federal Bureau of Investigation (FBI). We also used data from the National Center for State Courts to obtain information on misdemeanors and felonies~\cite{Nicole2020State}.

\subsection{Criminal-correlated socio-economic indicators}
To check the current socioeconomic environment in each state, we also collected 12 additional state-level features that are commonly referenced as the main correlates of crime rates. There are many statistics that are not collected on a yearly basis, and we therefore needed to find a way to obtain similar data. First, we gathered information on the economic status of each state. We collected the gross domestic product (GDP)~\cite{kyklos2010Does} for 2020 from the Bureau of Economic Analysis (BEA, \url{https://www.bea.gov}). We gathered personal income per capita (PCPI) data for 2020 from the BEA website~\cite{kyklos2010Does}. Happiness was collected from WalletHub \url{https://wallethub.com}, and is the level of happiness estimated from one’s socioeconomic status such as prices and unemployment rates~\cite{Nikolic2014Making}. We took the average expenditure of a single traveler in 2018 (travel spending) from the U.S. Travel Association (\url{https://travelanalytics.ustravel.org})~\cite{Boakye2010Studying}.

We also considered the demographic information for each state. We obtained the population density for large cities (Metropolitan Statistical Area (MSA)) in 2020 from the US Census Bureau (\url{https://data.census.gov/cedsci})~\cite{shelley1981crime,Watts1931influence}. An aging population is defined as the proportion of the population aged 65 years or older in 2020 ~\cite{Phillips2006relationship}. We also retrieved the proportion of the female population from the U.S. Census Bureau~\cite{Heidensohn1989Gender}. 

Government-related factors can also be candidates for correlates of crime rates. We obtained the total education expenditure for a single student (education) in 2018 from the Institute of Education Science (IES) report~\cite{2020Revenues}. In addition, we retrieved the average area covered by a single police officer (police area) in 2018 from the FBI (\url{https://crime-data-explorer.app.cloud.gov})~\cite{Lindstr2013More}. We use two different definitions of penalties for misdemeanors and felonies. A penalty is defined as the average among the maximum and minimum fines in 2019 for misdemeanors (\url{https://www.ncsl.org})~\cite{Tsebelis1990Penalty}. To account for the different solvencies for each state, we normalize this value by dividing the GDP per capita. However, we used the average among the maximum and minimum imprisonment periods in 2021 for felonies (\url{https://www.criminaldefenselawyer.com})~\cite{Hillsman1990Fines}. In addition, considering the average human lifespan, we regarded the death penalty or life imprisonment as 99 years of imprisonment.

In addition, our analysis incorporates the social atmosphere. We used the yearly averaged temperature level between 1971 and 2000 (i.e., temperature) retrieved from the NOAA National Clinical Data Center of the United States (\url{https://www.ncei.noaa.gov})~\cite{Field1992effect}, along with the percentage of gun owners in 2020 (gun ownership) taken from the RAND social and economic well-being (\url{https://www.rand.org})~\cite{Schell2020State-level}.

Although there are 50 states in the USA, only 30 have statistics for all 12 socioeconomic indicators, along with a criminal dataset. Therefore, we considered these 30 states based on their available data (see Table~\ref{tab:states}). 

\vspace{4mm}
\begin{table}
    \centering
    \begin{tabular}{ccccc}
    \noalign{\smallskip}\noalign{\smallskip}
    \hline\hline
    Northeast & Midwest & West & South \\
    \hline\hline
    Connecticut & Illinois & Arizona & Alabama \\
    \hline
    Maine & Iowa & California & Florida \\
    \hline
    Massachusetts & Kansas & Idaho & Kentucky \\
    \hline
    New Hampshire & Michigan & Nevada & Maryland \\
    \hline
    New Jersey & Minnesota & Utah & North Carolina \\
    \hline
    New York & Missouri & Washington & Texas \\
    \hline
    Pennsylvania & Nebraska & - & - \\
    \hline
    Rhode Island & Ohio & - & - \\
    \hline
    Vermont & Wisconsin & - & - \\
    \hline\hline
    \end{tabular}
    \vspace{1.5mm}
    \caption{List of selected states in the USA distinguished by their regional locations.}
    \label{tab:states}
\end{table}

\subsection{Selection of regression model}
In general, a regression is used to model the correlation between two or more variables. For a regression analysis, we tested two techniques: a linear regression~\cite{Seber2012Linear} and a logistic regression~\cite{Hilbe2009Logistic}. Although a simple regression is sufficiently powerful to account for the correlation between observations, it is difficult to expand to multiple independent variables. To find a more optimal method for the dataset, we estimated the coefficient of determination ($R^2$) for the two regression methods as follows: 

\begin{table}
    \centering
    \begin{tabular}{c|cccccc}
    \noalign{\smallskip}\noalign{\smallskip}\hline\hline
    & GDP & Penalty & Temperature & Education & Happiness & PCPI \\
    \hline
    \hline
    Linear & 0.05 & 0.40 & 0.77 & 0.77 & 0.92 & 0.89 \\
    \hline
    Logistic & -0.04 & -0.05 & -0.07 & 0.02 & -0.03 & 0.25 \\
    \hline
    \hline
    & Police area & Aging population & MSA & Gender & Travel spending & Gun ownership \\
    \hline
    \hline
    Linear & 0.68 & 0.89 & 0.93 & 0.93 & 0.50 & 0.82 \\
    \hline
    Logistic & -0.07 & -0.25 & -0.05 & -0.03 & -0.08 & -0.03 \\
    \hline
    \hline
    \end{tabular}
    \vspace{1.5mm}
    \caption{Coefficient of determination for linear and logistic regressions}
    \label{tab:table1}
\end{table}

To determine which method, logistic or linear, is more appropriate, Table~\ref{tab:table1} compares the $R^2$ values for each feature. Because the goal of the test was to determine the suitability of the two methods, we conducted it using misdemeanors. The target data were normalized to min-max scaling prior to applying the logistic regression. As indicated in the table, a linear regression shows a higher $R^2$ than a logistic regression. As a result, we chose a linear regression as the primary methodology for this analysis.

\subsection{Gradient-boosted trees}
Gradient boosting is a machine learning technique that is commonly used for regression and classification tasks. It generates a prediction model from an ensemble of many weak prediction models, most commonly, decision trees. When a decision tree serves as the weak model, the resulting algorithm is referred to as ``gradient-boosted trees,” which frequently outperforms a random forest. Although a gradient-boosted tree model is constructed in the same manner as other boosting methods, it differs in that it allows the optimization of any differentiable loss function. 

In this study, we used an optimized distributed gradient boosting library called XGBoost~\cite{Myles2004An}. We tested the feature importance of 12 collected variables to predict the crime rate by type (misdemeanors and felonies). We calculated the feature importance through a linear regression using the root mean squared error of XGBoost. We used conventional training/test splits with an 8:2 ratio and then trained the model using the following parameters: n\_estimators = 100, learning\_rate = 0.1, and max\_depth = 6.

\section{Results}

The goal of this study is to determine the similarities and differences between the two types of crimes (misdemeanors and felonies) as well as their correlates. Complex relationships exist between correlates and crimes that are unknown, and perhaps unconfirmed. To accomplish our primary objective, we consider simple correlations between certain correlates and crimes and then extend our analysis to account for multiple features simultaneously.

\subsection{Tendencies of misdemeanors and felonies}

Although there are numerous classification schemes for criminal activities, as stated in the Introduction, such a detailed classification varies by state. As a result, in this study, we compare two major types of crimes: misdemeanors and felonies. Numerous studies have concentrated exclusively on felonies, and misdemeanors have been frequently overlooked. However, we believe that both have a significant impact on quality of life; although misdemeanors do not cause serious harm to people, they may result in more serious crimes. As a result, our initial step was to compare the rates of these two types of crimes.

\vspace{4mm}
\begin{figure}[!ht]
    \centering
    \includegraphics[width=\textwidth]{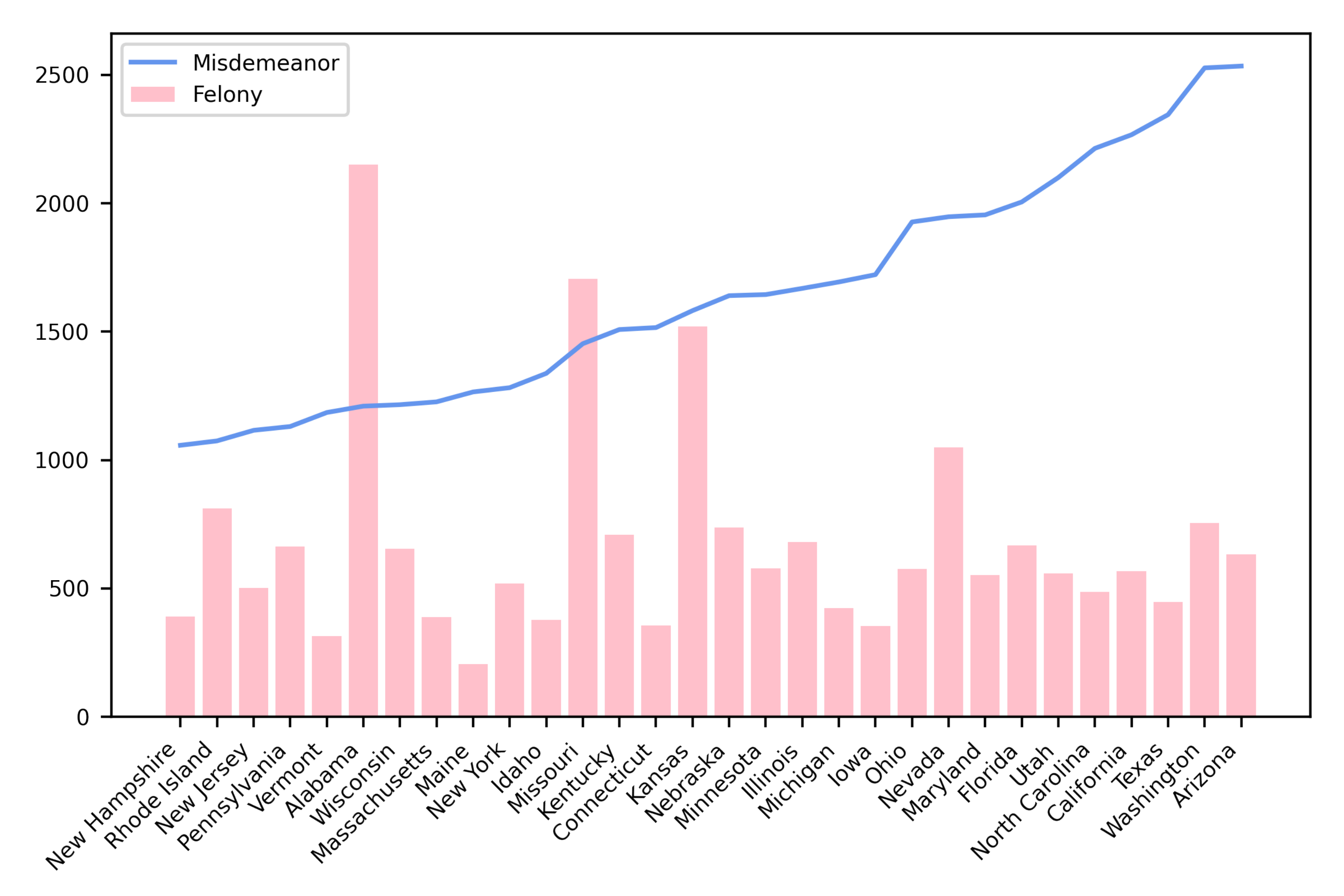}
    \caption{State-by-state breakdown of misdemeanors and felonies per $100,000$ population. The horizontal axis indicates the severity of the misdemeanors. A significant difference can be observed between the numbers of misdemeanors and felonies, indicating a low correlation of $r= -0.072$.}
    \label{fig:fig1}
\end{figure}

Figure~\ref{fig:fig1} illustrates the rates for two different types of crime per $100,000$ population. We arranged the states according to the magnitude of their misdemeanors (solid blue line). The rate of felonies (red bars) is significantly different from the rate of misdemeanors. In addition, we calculated the simple linear regression between the two rates, which yielded only $r = -0.072$. Thus, the rates of the two types of crimes are distinguishable; in other words, a state having a high rate of felonies does not necessarily have a high rate of misdemeanors. Thus, we analyzed the major correlates of misdemeanors and felonies separately in this study.

\subsection{Correlation of socioeconomic environment and crime rates}

Our finding of a difference in trend between misdemeanors and felonies persuaded us to investigate the underlying causative factors. Socioeconomic environments may have an effect on crime rates, but in different ways for the two types of crimes. As a result, we first conducted a regression analysis on the 12 frequently mentioned correlates (see the Methods section).

\vspace{4mm}
\begin{figure}[!ht]
    \centering
    \includegraphics[width=\textwidth]{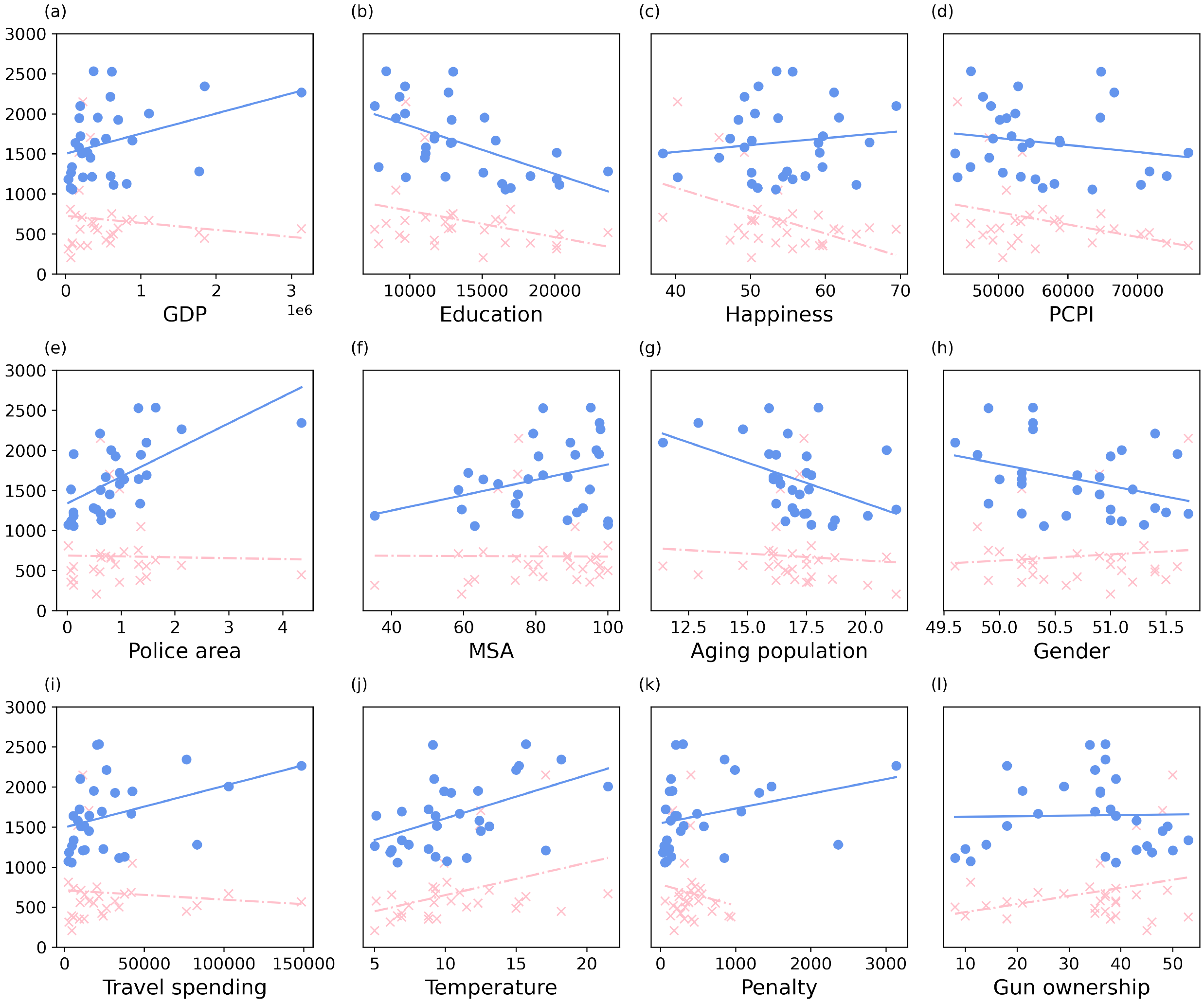}
    \caption{Analysis of linear regression between crime rates and 12 socio-economic characteristics of a state. Blue dots represent misdemeanor crime rates in each state, whereas coral dots represent felony crime rates. The solid lines in each panel represent the regression lines for misdemeanors (solid blue line) and felonies (solid red line), respectively. A significant difference was observed in the slope. Whereas all other characteristics of misdemeanors and felonies are identical, we calculated the penalties differently for the two types of crimes (see the Methods section).}
    \label{fig:fig2}
\end{figure}

\begin{figure}[!ht]
    \centering
    \includegraphics[width=\textwidth]{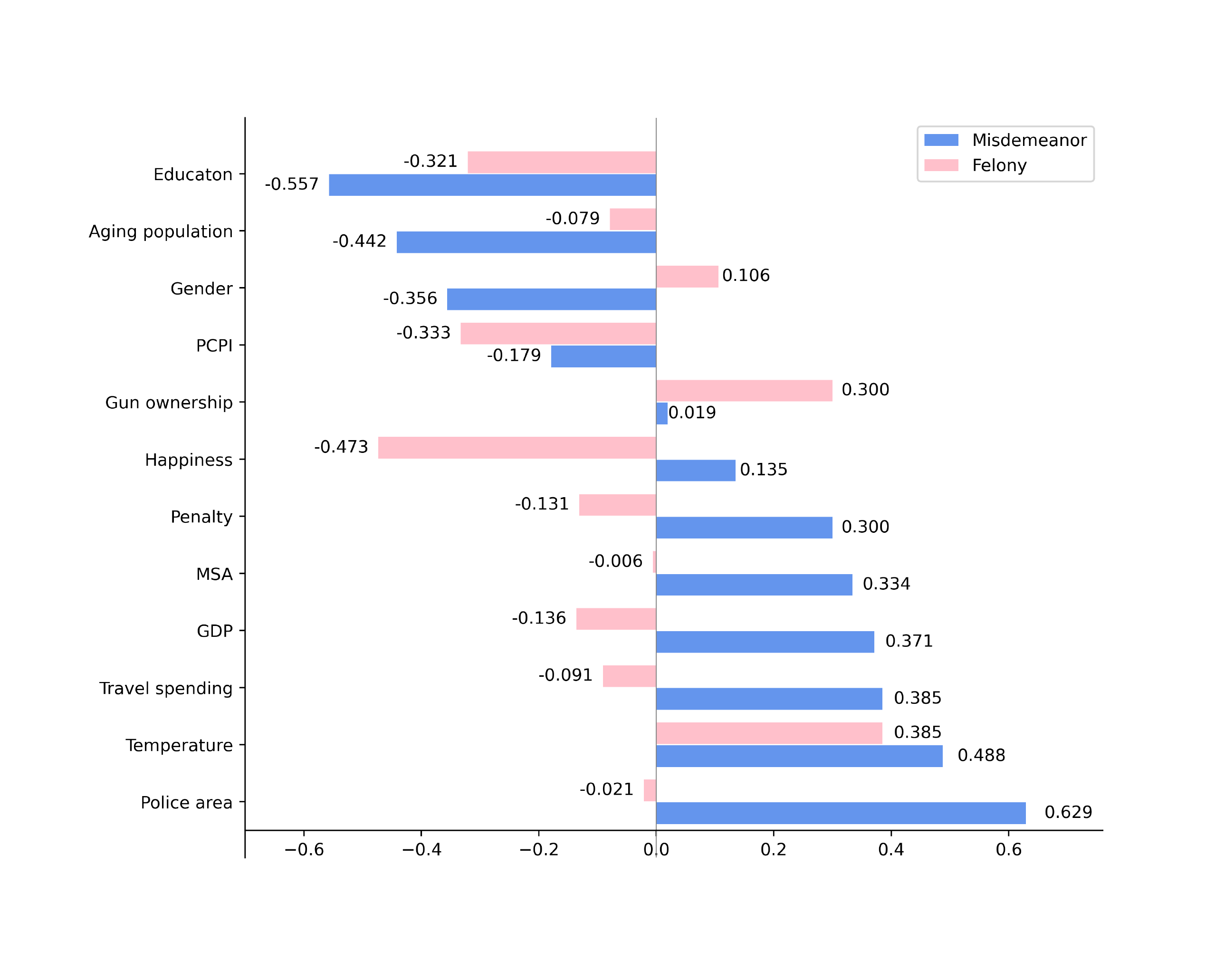}
    \caption{Pearson correlation between features and crime rates.}
    \label{fig:fig3}
\end{figure}

A linear regression analysis was conducted on each socioeconomic characteristic for a particular state~(Figures~\ref{fig:fig2}). The horizontal axis of each panel represents a specific socioeconomic indicator, and the vertical axis represents the crime rate. Whereas the signs of the regression slopes are identical in some cases, the regression lines for features such as GDP, happiness, travel spending, and penalties have opposite signs. In addition, the degree of inclination varied. For example, happiness has a significant negative correlation with felonies but no correlation with misdemeanors. By contrast, although police area appears to affect the number of misdemeanors, it appears to have no significant effect on the number of felonies. However, relying solely on Figure~\ref{fig:fig2} makes it difficult to determine which characteristics have a significant impact on each type of crime. To render a more detailed comparison, we computed the Pearson correlation between each feature and the crime rates (see Figure~\ref{fig:fig3}). 

\vspace{4mm}

\iffalse 
\begin{table}
    \centering
    \begin{tabular}{c|cccccc}
    \noalign{\smallskip}\noalign{\smallskip}\hline\hline
    & GDP & Penalty & Temperature & Education & Happiness & PCPI \\
    \hline
    \hline
    Misdemeanor & 0.37 & 0.30 & 0.48 & -0.55 & 0.13 & -0.17 \\
    \hline
    Felony & -0.13 & -0.13 & 0.38 & -0.32 & \textbf{-0.47} & -0.33 \\
    \hline
    \hline
    & Police area & Aging population & MSA & Gender & Travel spending & Gun ownership \\
    \hline
    \hline
    Misdemeanor & \textbf{0.62} & -0.44 & 0.33 & -0.35 & 0.38 & 0.01 \\
    \hline
    Felony & -0.03 & -0.07 & -0.01 & 0.10 & -0.09 & 0.30 \\
    \hline
    \hline
    \end{tabular}
    \vspace{1.5mm}
    \caption{Pearson correlation between features and crime rates.}
    \label{tab:table2}
\end{table}
\fi

When we disregard the signs, misdemeanors exhibit strong correlations in the following order: police area, education expenditure, and temperature. Meanwhile, felonies have significant (unsigned) correlations with happiness, temperature, and PCPI, among other variables. Thus, the primary correlates of misdemeanors and felonies are markedly different. Considering these signs, misdemeanors are reduced when a single police officer covers a smaller area, a state invests more in education, or the annual average temperature of a state is low. Similarly, a positive correlation exists between temperature and the number of felonies; however, the number of police officers per unit area has little effect on such number. Rather, happiness has the strongest negative correlation with the number of felonies, which could be explained by fear engendered by the high rate of major criminal activities~\cite{Hale1996Fear}.

In addition, we observed a negative correlation between the severity of a penalty and the number of felony convictions but a positive correlation between such severity and the number of misdemeanor convictions. Debates over retributive and restorative justice have been ongoing for decades~\cite{wenzel2008retributive}. Whereas some have argued that harsher punishments reduce crime rates, others have claimed that punishment should focus on reformation, allowing the convicted person to become a productive member of society. Our findings support the benefits of retributive justice, in part by demonstrating a negative correlation between the number of felonies and their associated penalties. However, our findings also suggest that this retributive standard is unsuitable for misdemeanors, as evidenced by the positive correlation between the penalty levels and number of misdemeanors. Indeed, one study discovered that fining individuals for minor rule violations can reduce the feeling of guilt and result in more violations~\cite{ariely2012honest}. The results of our study corroborate this finding. Thus, it can be concluded that misdemeanors and felonies should be treated differently; whereas more penalties may be beneficial in terms of reducing the number of felonies, one should reconsider this approach when regarding misdemeanors. We also found that gun possession is highly correlated with the felony rate; thus, it provides supporting evidence for the more-guns-cause-more crime hypothesis~\cite{duggan2001more}. 

\subsection{Other demographic factors and crime rates} \label{sub3}

The distinct patterns observed for felonies and misdemeanors raise an intriguing question: Do other demographic variables reflecting one's way of life have a correlation with crime? Political orientation is one possible correlate of the way of life and may affect the crime rate. For instance, political preferences in the USA can be estimated using the types of vehicles parked in parking lots~\cite{gebru2017using}. The USA has its own distinct presidential election procedure and can thus be classified into two groups: states that support the Republican Party and states that support the Democratic Party. We estimated the parties based on the results of the most recent presidential election in 2020~\cite{jerome2021state}. 

The correlation coefficients for misdemeanors and felonies by political orientation are shown in the upper and lower rows for Republicans and Democrats, respectively (Table~\ref{tab:table3}). Gun ownership has the highest correlation coefficient with misdemeanors in Republican-dominant states, whereas happiness is the most influential factor for felonies. By contrast, in the case of Democratic-dominant states, the highest correlation coefficients are for number of police officers and the number of those charged with misdemeanors whereas the proportion of elderly people is strongly correlated with the number of felonies. As a result, we discovered that the major correlates of crimes also differ based on the political orientation of the states.

\vspace{4mm}

\begin{table}
    \centering
    \begin{tabular}{c|c|cccccc}
    \noalign{\smallskip}\noalign{\smallskip}\hline\hline
    \multicolumn{2}{c|}{} & GDP & Penalty & Temperature & Education & Happiness & PCPI \\
    \hline
    \hline
    \multirow{3}{*}{Misdemeanor} & Republican & 0.51 & 0.56 & 0.24 & -0.19 & 0.42 & 0.44 \\
    & Democratic & 0.33 & 0.51 & 0.39 & -0.63 & 0.15 & -0.10 \\
    \hline
    \multirow{3}{*}{Felony} & Republican & -0.02 & -0.34 & 0.36 & 0.28 & \textbf{-0.69} & -0.04 \\
    & Democratic & 0.21 & 0.19 & 0.42 & -0.43 & -0.11 & -0.21 \\
    \hline
    \hline
    \multicolumn{2}{c|}{} & Police area & Aging population & MSA & Gender & Travel spending & Gun ownership \\
    \hline
    \hline
    \multirow{3}{*}{Misdemeanor} & Republican & 0.27 & -0.20 & 0.60 & -0.01 & 0.48 & \textbf{-0.86} \\
    & Democratic & \textbf{0.73} & -0.55 & 0.18 & -0.34 & 0.42 & -0.02 \\
    \hline
    \multirow{3}{*}{Felony} & Republican & -0.46 & 0.15 & -0.14 & 0.367 & -0.01 & 0.34 \\
    & Democratic & 0.36 & \textbf{-0.44} & -0.28 & -0.40 & 0.31 & -0.10 \\
    \hline
    \hline
    \end{tabular}
    \vspace{1.5mm}
    \caption{Pearson correlations between the socioeconomic factors and crime rates based on the political orientations of the states.}
    \label{tab:table3}
\end{table}

The follow-up question concerned the effect of geographic location on the crime rate correlates. Because the territorial expansion of the United States began on the east coast, culture varies by longitude. In addition, the northern states border Canada, whereas some of the southern states border Mexico, and as a result, the latter are heavily influenced by Mexican culture. The correlations between socioeconomic indicators and crime according to geographical location are shown in Table~\ref{tab:table4}. A complete list is presented in Table~\ref{tab:states}. In the northeast, travel spending is most strongly correlated with misdemeanors, whereas temperature is primarily associated with felonies. In the West, the highest correlation coefficients were found to be GDP for misdemeanors and education and happiness for felonies. The Midwest has the highest correlation coefficients for gun ownership and temperature for misdemeanors and felonies, respectively. Finally, in the South, GDP showed the highest correlation coefficient with misdemeanor convictions, felony convictions, population density in large cities, and elderly population density. However, because each category includes only seven states, and the differences are insignificant, it is difficult to draw a conclusion in a statistically significant manner.

\begin{table}
    \centering
    \begin{tabular}{c|c|cccccc}
    \noalign{\smallskip}\noalign{\smallskip}\hline\hline
    \multicolumn{2}{c|}{} & GDP & Penalty & Temperature & Education & Happiness & PCPI \\
    \hline
    \hline
    \multirow{3}{*}{Misdemeanor} & Northeast & 0.22 & 0.30 & -0.21 & 0.26 & 0.18 & 0.36 \\
    & West & 0.77 & \textbf{0.82} & 0.60 & 0.42 & -0.37 & 0.31 \\
    & Midwest & 0.45 & 0.41 & -0.11 & 0.36 & 0.04 & -0.09 \\
    & South & \textbf{0.88} & 0.71 & 0.37 & -0.60 & 0.54 & 0.54 \\
    \hline
    \multirow{3}{*}{Felony} & Northeast & 0.43 & 0.11 & \textbf{0.63} & 0.09 & -0.13 & 0.13 \\
    & West & 0.25 & -0.54 & 0.37 & \textbf{0.65} & \textbf{-0.65} & 0.48\\
    & Midwest & -0.36 & -0.04 & \textbf{0.65} & -0.26 & -0.30 & 0.11 \\
    & South & \textbf{-0.77} & -0.28 & -0.08 & 0.37 & -0.65 & -0.65 \\
    \hline
    \hline
    \multicolumn{2}{c|}{} & Police area & Aging population & MSA & Gender & Travel spending & Gun ownership \\
    \hline
    \hline
    \multirow{3}{*}{Misdemeanor} & Northeast & 0.28 & -0.08 & -0.05 & \textbf{0.42} & 0.39 & 0.08 \\
    & West & 0.31 & 0.11 & 0.48 & 0.55 & 0.37 & -0.48 \\
    & Midwest & 0.30 & 0.27 & 0.28 & 0.37 & 0.46 & \textbf{-0.81} \\
    & South & 0.31 & -0.54 & 0.71 & -0.60 & 0.77 & -0.42 \\
    \hline
    \multirow{3}{*}{Felony} & Northeast & -0.21 & -0.41 & 0.61 & 0.30 & 0.51 & -0.53 \\
    & West & -0.14 & 0.28 & 0.31 & 0.00 & 0.60 & -0.06 \\
    & Midwest & -0.48 & -0.60 & -0.09 & 0.06 & -0.30 & 0.62 \\
    & South & -0.08 & \textbf{0.77} & \textbf{-0.77} & -0.48 & -0.60 & 0.48 \\    
    \hline
    \hline
    \end{tabular}
    \vspace{1.5mm}
    \caption{Pearson correlations between the socioeconomic factors and crime rates by the geographical location of the states. Values in bold show the most significant correlations among the features.}
    \label{tab:table4}
\end{table}

As can be seen, in the USA, different characteristics affect crime rates based on both regional and political orientations. However, the results presented in this section are not statistically significant.

\subsection{Multi-factor analysis for the crime rate}

Thus far, we have examined the factors that influence crime rates. Our identification of multiple factors raises two critical questions: Is the crime rate truly related to our correlates, and is there a single socioeconomic indicator that is more significant than the others? The first question can be answered based on the high statistical significance indicated by the large correlation coefficients described in the preceding sections. However, the preceding analysis was conducted under the assumption that the correlates are unrelated, which is not the case in reality. Consequently, when all factors are considered concurrently, previous findings do not accurately reflect the true influence of each factor. To address the second question, we used XGBoost to conduct a regression analysis of the crime rate (see the Methods section). We compared the importance of each feature based on the type of crime. We did not apply hyperparameter tuning because the purpose of this analysis was to compare the relative impact of the features rather than to make absolute predictions.

\begin{figure}[!ht]
    \centering
    \includegraphics[width=\textwidth]{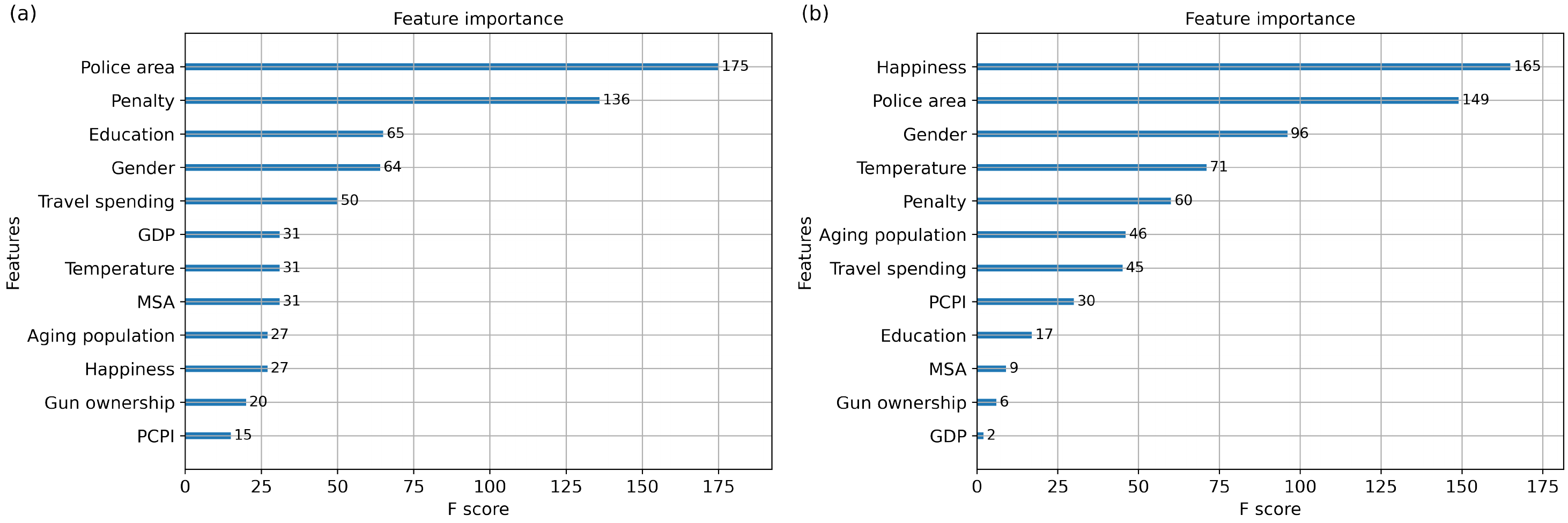}
    \caption{Feature importance score for (a) misdemeanors and (b) felonies.}
    \label{fig:fig4}
\end{figure}

Figure~\ref{fig:fig4} illustrates the feature importance of the results of turning a simple model for two crimes using XGBoost. This feature importance does not consider the sign of the influence (positive or negative) but rather the degree of influence. In the case of misdemeanors (Figure~\ref{fig:fig4} (a)), the critical factors are the number of police officers per unit area, education investment, and penalties. Thus, to a certain degree, the result is similar to the previous analysis using a single feature (see Figure~\ref{fig:fig3}). For felonies (Figure~\ref{fig:fig4} (b)), the following factors show high importance scores: happiness, the number of police officers per unit area, ratio of gender, and the average temperature. When considering only a single feature, felonies demonstrated the importance of happiness, PCPI, and temperature. In other words, although the results are not identical to those obtained from the single-feature analysis, they are comparable (see Figure~\ref{fig:fig3}). In other words, we confirmed that the analysis applied using one feature dataset per target dataset was consistent with the multi-factor analysis to a certain degree, enhancing its reliability. 

\section{Discussion}
In this study, to aid in crime reduction, we attempted to identify those factors that have a significant impact on crime rates. In other words, the analysis was conducted under the assumption that identifying and addressing the root causes of crime will naturally result in a crime reduction. Our findings suggest that we should not treat all crimes identically because the natures of felonies and misdemeanors differ. 

After separating the two types of crimes, we found that misdemeanors are significantly influenced by the number of police officers per unit area and educational expenditure. Thus, the solution may include increased police deployment per unit area and increased educational funding. Compared to misdemeanors, felonies were significantly influenced by happiness and temperature in that order. To decrease the rate of felonies, happiness should be quantified using factors such as product prices and unemployment rates; thus, various economic policies should be implemented. As illustrated in Figure~\ref{fig:fig3}, the characteristics with inverse correlation coefficients between misdemeanors and felonies, such as happiness, gun ownership, and GDP, should be improved with caution because changes may have a positive effect on the number of felonies but a negative effect on the number of misdemeanors. Particular care should be taken regarding the penalty because it has a positive correlation with misdemeanors, despite potentially aiding in a decrease in the felony rate.

Owing to limited availability, the data analyzed in this study focused on 30 states in the USA. Additional data may be required to achieve a higher statistical significance, and a follow-up study is required to conduct a more objective analysis. In addition, because only data from the USA were used, additional research is necessary to determine whether the findings of this study apply to other countries and whether crime rates will decrease as a result of policies that reflect the findings. However, we hope that further studies considering misdemeanors are warranted to enhance the quality of life of citizens.

\section{Acknowledgements}
We thank Dr. Taekho You for the valuable and constructive suggestions. This research received an institutional grant from Soongsil University. The Korea Institute of Science and Technology Information (KISTI) also supported this research by providing KREONET, a high-speed Internet connection. The funders had no role in the study design, data collection and analysis, decision to publish, or manuscript preparation. 

\section{Author contributions}
All authors designed the research and wrote the manuscript. Juyoung Kim collected and analyzed the data.

\bibliographystyle{unsrt}
\bibliography{biblio} 
\end{document}